\documentclass{elsart}
\usepackage{graphicx}
\def\contrac#1#2#3<#4>{%
\setbox1=\hbox{$#1$}%
\setbox2=\hbox{$#2$}%
\setbox3=\hbox{$#3$}%
\dimen0=.5\wd1%
\advance\dimen0 by \wd2%
\advance\dimen0 by .5\wd3%
\setbox0=\hbox{\kern.5\wd1%
\vtop{\hbox to \dimen0{\vrule depth#4\hfil\vrule}\hrule}}%
\dimen0=#4 \advance\dimen0 by \lineskip \dp0=\dimen0%
\wd0=0pt%
\setbox1=\vtop{\lineskiplimit=10cm\lineskip=1mm \box1\box0}%
\box1\box2\box3}

\begin{document}
\begin{frontmatter}
\title{\Large \bf Mean-field Based Approaches to  Pairing Correlations
in Atomic Nuclei}
\author{M. Anguiano\thanksref{marta}},
\author{J.L. Egido},
\author{L.M. Robledo}
\address{Departamento de F\'{\i}sica
Te\'orica, Universidad Aut\'onoma de Madrid, E-28049 Madrid, Spain} 
\date{\today}
\thanks[marta]{Present address: Dipartamento di Fisica, Universita di Lecce,
73100 Lecce, Italy.}
\begin{abstract}
The evolution of the pairing correlations from closed shell
to midshell nuclei is analyzed  in the Sn isotopes with the Finite Range Density 
 Dependent  Gogny force. As theoretical
approaches we use the Hartree-Fock-Bogoliubov, the Lipkin-Nogami, 
their particle number projected counterparts and the full 
variation after particle number projection method. We find that whereas 
all approaches succeed rather well in the description of the total energy 
they differ significantly in the pairing correlation content of the wave 
functions. The description of the evolution from the weak to the strong pairing 
regime  is also approach dependent, specially at shell closure.

\end{abstract}
\end{frontmatter}

%
%

   Pairing correlations play an important role in the understanding of
nuclear phenomena. Observables like moments of inertia, level densities
and energies of the lowest-lying excited states, to mention a few, are
strongly influenced by  these correlations. In spite of their
relevance pairing correlations are still not well understood. On the
experimental side, because the pairing energy itself is not an observable, 
it is not easy to extract relevant information from the  
data (at high angular momentum, for example, we have gapless 
superconductivity). On the theoretical side, the basic problem is 
 the mean field approximation to the pairing field which is not as 
 effective as it is with the deformation (Hartree-Fock) field. 

Recently, new experimental techniques  have made possible
to access regions far away from the stability line allowing to study, 
among others, very exotic systems where pairing correlations play a key
role to make the nuclei bound \cite{BE.91}. The experimental studies 
of N=Z nuclei also provide useful information on the
proton-neutron pairing, see  for example reference \cite{RU.96}. 
It seems therefore timely 
to study  the pairing correlations with a force able to provide both binding
and pairing energies.
  In the past there have been several theoretical approaches to the nuclear 
pairing problem most of them using schematic or separable forces with little
predictive power. From a more fundamental point of view effective forces
should be used in these studies. The most popular of the density dependent 
forces, the Skyrme forces, in general use a different force 
 for the particle-particle  than for the particle-hole channel
and so do, in general, the relativistic approaches. The only  density dependent
 force which has
 a selfcontained pairing force, because of its finite range, is the Gogny 
force \cite{Gog80}.  This property of the
Gogny force makes it unique to study different theoretical approaches because
the renormalization of the force is, in principle, not needed due to its 
selfcontainedness. The purpose of this letter is to investigate the pairing correlations
in different mean-field based approaches along different pairing regimes
using  the finite range density dependent Gogny interaction.

The simplest microscopic approach to describe the nuclear many-body system is
 the Hartree-Fock (HF) theory. The HF wave function is an antisymmetrized 
 product of single 
particle wave functions determined in a variational way. The particles move
independently in the HF orbitals determined selfconsistently excluding 
thereby any particle-particle correlation besides the ones considered
in the common mean-field potential.
   The basic and oldest approach to include  particle-particle
correlations, i.e., pairing correlations, is the
BCS approach \cite{BCS57}. This is still a mean field approach, where the
wave function is a product of quasiparticle operators
$\beta_k$, i.e., \( | {\rm BCS} \rangle \propto \prod \beta_k |-\rangle \),
given by the Bogoliubov-Valatin transformation
\begin{equation}
\label{eq:BVtra}
 \beta^{\dagger}_k = u_k a^{\dagger}_{k} - v_k a_{\bar{k}}, \quad
  \beta^{\dagger}_{\bar{k}} = u_k a^{\dagger}_{\bar{k}} + v_k a_{k}
\end{equation}
with  $a_k, a^{\dagger}_{k}$ the annihilation and creator particle operators
in the Hartree-Fock basis and $\bar{k}$ the time reversal orbital to $k$. 
The coefficients $u_{k},v_{k}$ are determined by 
the Ritz variational principle.
Since the Ansatz of Eq.~(\ref{eq:BVtra}) mixes creation and annihilation operators 
the wave function $| {\rm BCS} \rangle$ is not an eigenstate of the particle number
operator. The variational equation is therefore
\begin{equation}
 \frac{\delta}{\delta \Phi}\langle \Phi |\hat{H} |  \Phi\rangle
 - \lambda \frac{\delta}{\delta \Phi}\langle \Phi |\hat{N}|  \Phi\rangle = 0,
\label{eq:minmfa}
\end{equation}
with $|\Phi \rangle =| {\rm BCS }\rangle$ and $\lambda$ the Lagrange multiplier 
determined under the constraint that 
the BCS wave function has on the average the  particle number N. 
 In spite of its simplicity and success the BCS approach lacks selfconsistency
in the sense that the HF  and the pairing fields are not treated on the 
same footing, i.e., first the HF orbitals are calculated and then their occupancies
around the Fermi surface determined by the BCS equations. The theory which remedies
this drawback is the Hartree-Fock-Bogoliubov (HFB) approach. This is again
 a mean field 
approach, with wave function \( | {\rm HFB} \rangle = \prod \alpha_k |-\rangle \), 
and with quasiparticle operators $\alpha_k$ determined by the generalized Bogoliubov
transformation
\begin{equation}
\label{eq:Bogtra}
\alpha_k = \sum U^*_{lk} c_l + V^*_{lk} c^{\dagger}_l.
\end{equation}
where  $c_l, c^{\dagger}_{l}$ are the annihilation and creator particle operators
in a suitable  basis.
In this case the variational parameters are the matrices $U, V$ which
are determined by minimization of Eq.~(\ref{eq:minmfa}) but now with 
$| \Phi \rangle =| { \rm HFB }\rangle$. The mean field approaches
(HF, BCS, HFB) have been widely used over the years with simple separable
forces, effective forces and relativistic ones, to describe many nuclear 
properties and provide the backbone to theories beyond mean field as the 
Random-Phase-Approximation (RPA) or the Generator Coordinate Method (GCM).

The success of the mean field approaches  is based on their  ability 
to deal with single particle
motion as well as with the collective motion associated with
symmetries~\footnote{ Continuous symmetries, as rotations
in any space:  coordinate space, gauge space of particle number 
operator, etc  as well as discrete symmetries, e.g. spatial parity.}. 
The collective degrees of
freedom are incorporated in the variational Hilbert space by the 
spontaneous symmetry breaking mechanism. The wave functions of this
enlarged Hilbert space  are not eigenstates of the symmetry operators
and  are usually constrained to obey the symmetries on the average.
For most symmetries the mean field approach is  very satisfactory, 
for instance, for the rotational motion associated to the angular momentum,
 see  Ref.~\cite{RS.80} for a thorough discussion.
In the case of pairing correlations, in which we are interested in this letter,
the crucial quantities are  the number of correlated 
pairs and the level density around the Fermi surface. If these quantities are  
small, and in nuclei they usually are, mean field theories are not enough and
one should do something better.

The semi-classic recipe of solving the BCS and HFB equations with a constraint 
on the particle number operator can be derived as the first order result of a 
full quantum-mechanical expansion (the Kamlah expansion) \cite{KAM.68} of
the particle number projected quantities in terms of unprojected ones.
 The second order in this
expansion takes into account  the particle number fluctuations and might
cure some of the deficiencies of the first order approximation. However,
full calculations up to second order are rather cumbersome 
\cite{ZSF.92,FO.97,VER.01}
and most second order calculations have been done using the Lipkin-Nogami
(LN) recipe  proposed in Refs.~\cite{LIP.60,NOG.64,GN.66}.
The original formulation of the LN method was for a
simple separable pairing interaction but it has been  recently extended to 
non-separable ones  \cite{RNB.66} and  density
dependent finite range forces \cite{VER.97,VER.00}. 
The variational equations of the LN method are given by

\begin{equation}
 \frac{\delta}{\delta \Phi}\langle \hat{H} - h_2 (\Delta\hat{N})^{2}\rangle
 - h_1 \frac{\delta}{\delta \Phi}\langle \hat{N}\rangle = 0,
\label{eq:ln}
\end{equation}
with $h_1$ and $h_2$ given by
\begin{eqnarray}
  h_{1} & = & \frac{\langle\hat{H}\Delta\hat{N}\rangle - 
  h_{2} \;\langle(\Delta\hat{N})^{3}\rangle}
 {\langle(\Delta\hat{N})^{2}\rangle}  \label{h1}\\
  h_{2} & = & \frac{\langle(\hat{H}-\langle\hat{H}\rangle)
  (\Delta\hat{N})^{2}\rangle  -
 \langle\hat{H}\Delta\hat{N}\rangle\langle(\Delta\hat{N})^{3}\rangle/
 \langle(\Delta\hat{N})^{2}\rangle}
 {\langle(\Delta\hat{N})^{4}\rangle - \langle(\Delta\hat{N})^{2}
 \rangle^{2}- \langle(\Delta\hat{N})^{3}\rangle^{2}/\langle
 (\Delta\hat{N})^{2}\rangle}.
\end{eqnarray}
with  \(\Delta \hat{N} = \hat{N}- \langle \hat{N} \rangle \) and 
$\langle \hat{O}\rangle \equiv \langle \Phi |\hat{O}| \Phi\rangle$
for any operator $ \hat{O}$.
In the LN approach the $h_2$ parameter
is not varied, contrary to what a variational method would require,
 but only updated in each iteration of the minimization
process. In this respect the LN approach is not a fully variational method and
 its success  not quite well understood. 

In a mean field based approach, the ideal treatment of pairing correlations 
in nuclei is particle 
number projection (PNP) before the variation  \cite{DMP64}. This theory 
is rather complicated and up to now has  been mainly applied to  
separable forces \cite{EGPNP} or in small configuration spaces \cite{Carlo}.
 Only recently \cite{AER.01P} an exact 
particle number projection has been performed with finite range forces,
the Gogny forces, and large configuration spaces.   Let $|\Phi \rangle$ be a 
product wave function of the HFB type, i.e. a particle 
number symmetry violating wave function. We can generate
an eigenstate $|\Psi_{N}\rangle$ of the particle number  
by the projection technique \cite{RS.80}
\begin{equation}
   |\Psi_{N}\rangle = \hat{P}^{N}|\Phi \rangle= 
   \frac{1}{2\pi} \int_{0}^{2\pi}d\phi
e^{i(\hat{N}- N )\phi}  |\Phi \rangle.
\end{equation}
The particle number projected energy is given by
\begin{eqnarray}
 E^{N}_{proj}  =  \frac{\langle\Psi_{N}|\hat{H}|\Psi_{N} \rangle}
  {\langle\Psi_{N}|\Psi_{N} \rangle}=
    \frac{
   \int_{0}^{2\pi}d\phi e^{-i\phi N}\langle\Phi|\hat{H}e^{i\phi\hat{N}}
   |\Phi\rangle}
   {\int_{0}^{2\pi}d\phi e^{-i\phi N}\langle\Phi|
   e^{i\phi\hat{N}} |\Phi\rangle} = \int_{0}^{2\pi}d\phi \, y(\phi) E(\phi)
 \label{bigeq}
\end{eqnarray}
with
\begin{equation}
y(\phi) = \frac{e^{i\phi(\hat{N}-N)}} {\int_{0}^{2\pi} d\phi \langle\Phi|
   e^{i\phi(\hat{N}-N)} |\Phi\rangle}, \,\,\,\,\,\,\,
   E(\phi) = \frac{\langle\Phi|\hat{H}e^{i\phi\hat{N}}
   |\Phi\rangle}{\langle\Phi|e^{i\phi\hat{N}}
   |\Phi\rangle}.
 \label{yephi}
\end{equation}
One should distinguish  the projection after variation method (PAV) from
the variation after projection method (VAP). In the PAV approach
the wave function $|\Phi\rangle$ is determined by solving 
Eqs.~(\ref{eq:minmfa}) with  $| \Phi \rangle $ a HFB wave function and with
this wave function one can calculate the projected energy with
Eq.~(\ref{bigeq}).  
 In the VAP approach the wave function $|\Phi\rangle$ is determined minimizing
the projected energy, Eq.~(\ref{bigeq}). Obviously the VAP method provides
a much better approach. In our case all the variational equations are solved
 by the Conjugate Gradient Method \cite{ELM.95}. 

 As mentioned above we want to investigate the pairing correlations
in different mean-field based approaches using  the Gogny interaction
in the numerical applications. For this purpose we investigate some 
properties of the Sn isotopes ($Z=50$) from the $N=50$ shell closure 
to the $N=82$ one. The aim is to study the evolution of the pairing 
characteristics from the weak pairing regime (around the shell closure)
to the strong pairing  regime (midshell) to investigate
the quality of the different approaches. We have studied the ground state 
properties of the Tin isotopes in the ${\rm HF}$,  ${\rm HFB}$, and  ${\rm LN}$
approaches.  We have also performed projected calculations of the  
`after variation' type, the PAV and the PLN (Projected LN). 
In the PAV (PLN) method the intrinsic
wave function \( | \Phi \rangle \) is determined by the HFB (LN) equations,
afterwards a particle number projection on this wave function is performed
allowing the calculation of projected expectation values. Furthermore we have 
performed the VAP calculation. We think that this comparison is important
because the studies performed so far were done with separable forces, mostly
the monopole pairing.
The calculations have been performed with a triaxial code and with N$_0$=11 
oscillator shells. Furthermore the D1 parametrization of the Gogny force
has been used \cite{Gog80}.
To prevent the appearance of divergences associated with the neglection of
the exchange terms in the particle number projected calculations
\cite{AER.01P}, all calculations have been performed including {\it all} 
exchange terms of the forces, and all terms have been calculated without
any approximation \cite{AER.01}. In the density dependent term of the Gogny
 force we have used the projected density prescription, see  \cite{AER.01P}
  for more details.

   To investigate the pairing correlations we should look for a quantity
which can be defined in all approaches. In BCS theory with monopole pairing
usually the gap parameter has been  used. This quantity is strongly
related with the particle-particle correlation energy used in 
 the HFB approach, 
\begin{equation}
E_{pp}~=~- \frac{1}{2}  {\displaystyle Tr}  \left ( \Delta {\kappa}^*
\right),
\label{eq:enepp}
\end{equation}
with \({\Delta}_{kl}  = \frac{1}{2} \sum_{mn} 
{\bar{v}}_{klmn}{\kappa}_{mn} \) and
 \(\kappa_{mn} = \langle \Phi |c_n c_m | \Phi \rangle \) and $v$, in our
case, the Gogny and Coulomb interactions. We would like to remind that the
density dependent part of the Gogny force does not contribute to the pairing
field. In the LN approach we keep the same
definition, that means, the contribution of the  $- h_2 <(\Delta \hat{N})^2>$
term to $E_{pp}$ is not considered, this contribution is added to the HF energy.
The equivalent expression in the PNP case is 
( {cf} Eqs.(\ref{bigeq}-\ref{yephi}))~:
\begin{equation}
E_{pp}~=~ \int_{0}^{2\pi}d\phi \, y(\phi) E_{pp}(\phi)=
- \frac{1}{2} \int_{0}^{2\pi}d\phi \, y(\phi)
 {\displaystyle Tr}  \left ( \Delta^{10}(\phi) {\kappa}^{01}(\phi) \right)
\label{eq:eneppp}
\end{equation}
with \( {\Delta}^{10}_{kl} (\phi) = \frac{1}{2} \sum_{mn} 
{\bar{v}}_{klmn}{\kappa}^{10}_{mn} (\phi) \), and 

\begin{equation}
{\kappa}_{kl}^{10} (\varphi) =  \frac{ \langle \Phi | c_{l}c_k 
e^{i\varphi\hat{N}} | \Phi \rangle }{\langle \Phi | 
e^{i\varphi\hat{N}} | \Phi \rangle }, \,\,
{\kappa}_{kl}^{01} (\varphi) = 
\frac{ \langle \Phi | c_{k}^{\dagger}
c_{l}^{\dagger} e^{ i\varphi\hat{N}} | \Phi \rangle }{\langle \Phi | \
e^{i\varphi\hat{N}} | \Phi \rangle }. 
\end{equation}


\begin{figure}[h]
\begin{center}
\parbox[c]{14cm}{\includegraphics[width=14cm]{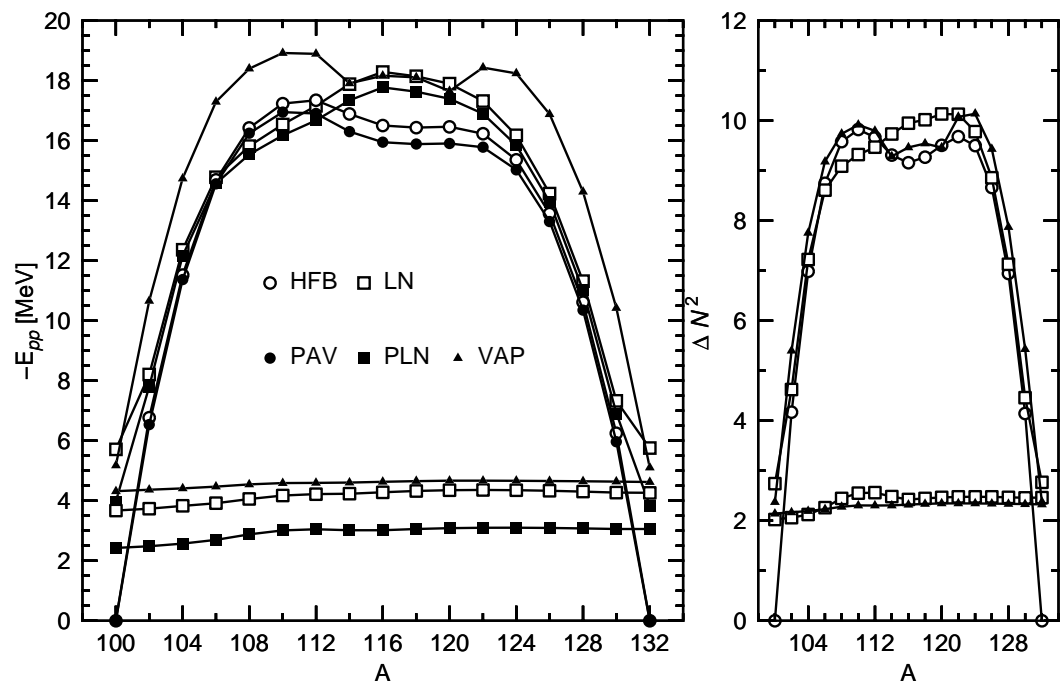}}
\end{center}
\caption{ Left panel: The particle-particle correlation energy \( E_{pp}\)
for the Sn isotopes in different approaches, for neutrons (protons) on the
top (bottom).
 Right panel: The particle number
fluctuation in the HFB, LN and the intrinsic wave function of the VAP 
approaches. Symbols and conventions are given in the left panel.}
\label{Fig1}
\end{figure}

In the left panel of Fig.~\ref{Fig1} we present $E_{pp}$ for the Tin isotopes
 in the
five approaches  mentioned above, calculated via Eq.~(\ref{eq:enepp}) for 
non-projected theories
and Eq.~(\ref{eq:eneppp}) for projected theories. Curves on the top (bottom) 
of the left panel 
correspond to  the neutron (proton ) pairing correlations.
The common wisdom about the general behavior of pairing correlations
along a  shell 
is that they behave as a semicircle~: at shell closure there are not pairing 
correlations,
 as we keep adding particles the pairing correlations increase 
until  midshell where they reach the maximum value.  From this point on they 
decrease up to the point where the next shell closure is reached.
This wisdom is mainly based on the mean field approach (BCS and HFB). As
a matter of fact the neutron pairing energy  in the  ${\rm HFB}$
approach  resembles very much this pattern. 
{\it Qualitatively } the behavior
of the neutron ${ E_{pp}}$ in the five approximations is similar~: they increase 
 from both shell closures towards midshell, up to N=58 and N=74 
where a kind of plateau develops from A=110 up to 122.
 {\it Quantitatively}, however, we find three distinctive behaviors corresponding 
 to the three
different intrinsic wave functions, the HFB and PAV approaches, the LN and
PLN and finally the VAP one. Both, HFB and PAV, give no pairing
correlations at the shell closures and display a kind of two hump behavior
as a function of the mass number.
The LN and PLN approaches, on the other hand, provide non-zero correlation
energy at the shell closures (the LN overestimate the VAP result) and display
a one hump behavior. The VAP approach, finally, provides the largest correlation
energies, showing an even more pronounced two hump structure, and giving non-zero 
correlations at the shell closures. 

If we exclude the VAP results we find a clear
trend in the other  approximations as a function of the mass number~:
At the shell closures and in their nearest vicinity there are large differences
in the results for the different approaches.  As we move to the middle of the
major shell the four approximations provide results rather close to each other.
Finally, around the middle of the major shell the dispersion in the results 
of the different approaches get larger. This peculiar behavior might be related,
as we shall see later, with the number of subshells involved in the pairing 
mechanism.
A remarkable point is the behavior of the LN (and the PLN to a lesser extend)
neutron correlation energies at the shell closures. We observe that the value 
of ${ E_{pp}}$ at the shell closures, i.e. N=50 and N=82, is not the 
extrapolation of the neighboring values, as it is the case for the HFB, PAV and 
VAP approaches. As we shall see later on this behavior is probably related with a 
degradation of the LN solution.
  
  The proton correlation energies are depicted in the lower part of the same
panel. The HFB and PAV energies, as expected  are zero for all N values (we
have a major shell closure at Z=50) while the LN, PLN and VAP are
not. They are not as large as the neutron ones, but the important point is
that they are different from zero, i.e., these theories are  able to 
gather a certain correlation energy from the so-called dynamical pairing.
Interestingly the projected versions of the HFB (for neutrons) 
and LN (for neutrons and protons) approaches provide smaller particle-particle 
correlations  than the unprojected ones. 
 We also find that the reduction produced in the LN case for protons is large
 (about 30 $\%$ )
as compared with the one for neutrons or the HFB case (5 $\%$ ). 
This result seems to
indicate that the LN approach in the closed shell regime overestimates the
pairing correlations (cf. \cite{DN.93}).

  A quantity that plays an important role in the 
approximate particle number projection methods is the fluctuation of the
particle number operator.  This quantity provides a measure on the degree of
symmetry breaking in the corresponding wave function. This obviously only
applies to symmetry breaking wave functions like BCS or HFB but not to 
symmetry conserving ones like HF or PNP wave functions.
 HFB or BCS wave functions with large \( <(\Delta \hat{N} )^2> \)
are expected to have large pairing (gauge deformation) correlation energies. 
   In the right panel of Fig.~\ref{Fig1} the fluctuations of the
particle number operator are shown.  In the projected approaches the intrinsic
wave functions are used in the calculations, consequently only the HFB, LN,
and VAP approaches are shown. Symbols and conventions are the same as in the 
left panel. As expected the overall behavior is similar to the  $E_{pp}$
in the corresponding approaches. It is interesting that, though the intrinsic
wave functions of the VAP and PAV approaches
have similar ``deformation'', in the gauge space associated to the particle number
operator, for many isotopes, the projection 
is much more effective producing larger ${ E_{pp}}$ in the case of the VAP than
in the PAV one. 

\begin{figure}[h]
\begin{center}
\parbox[c]{10cm}{\includegraphics[width=8cm]{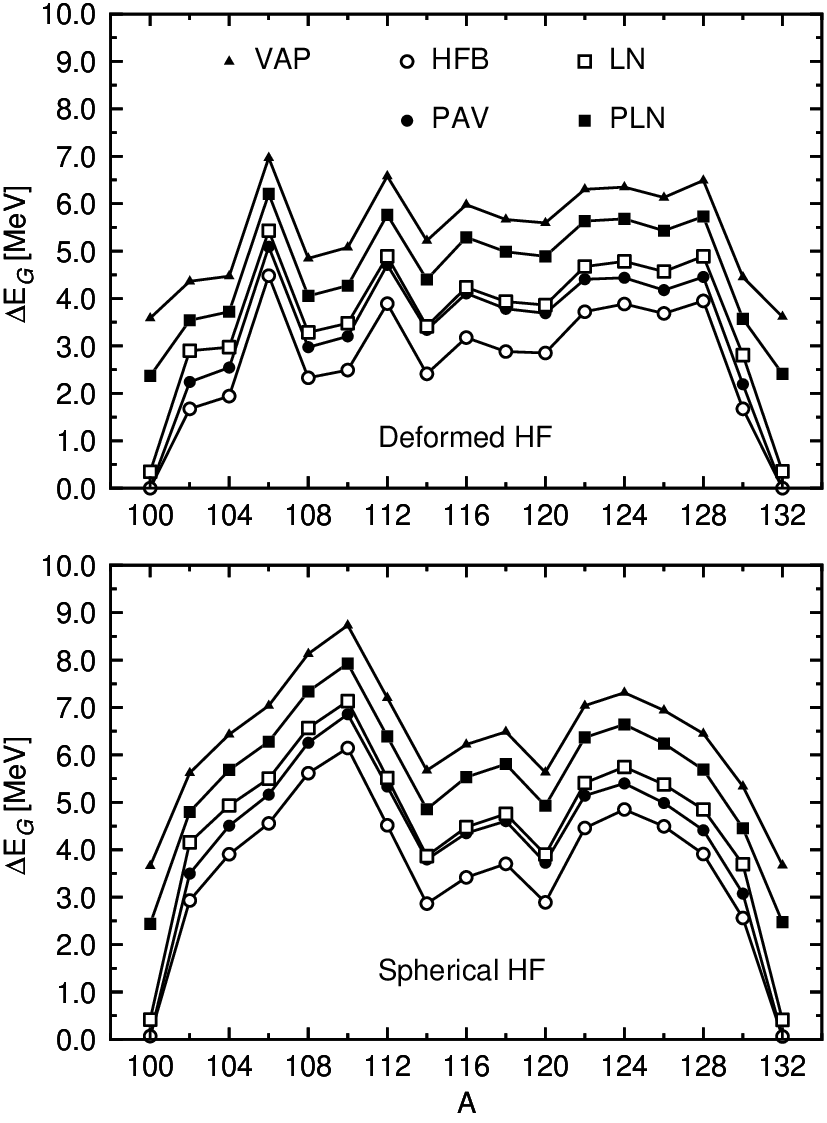}}
\end{center}
\caption{Upper part: The energy difference between the binding energy in the 
HF approach and the other theoretical approaches as a function of the
mass number. Lower part: Same as upper part but constraining the HF solutions
to spherical shapes.}
\label{Fig2}
\end{figure}

Another way to measure the pairing correlations in a given approach 
is to look for the energy gain obtained  by going from the HF (no pairing
correlations allowed) to the respective approach (pairing correlations allowed). 
In this way we may define
\begin{equation}
\Delta E_G = E_{HF} - E_{approach}
\label{eq:enepc}
\end{equation}
where $E_{HF}$ and $E_{approach}$ are the total binding energies in the HF 
theory 
and in  the corresponding approach. In this case ${approach}$ stands for 
any of the HFB, PAV, LN, PLN and VAP approaches. Obviously, $ \Delta E_G$ is 
 a measure of the energy gain when particle-particle correlations are allowed.

In the upper part of Fig.~\ref{Fig2} we display $ \Delta E_G$  for the
different approximations. The largest energy gain is provided by the VAP
method followed by the PLN one. Interestingly, at variance with
the $E_{pp}$ case,  the PAV approach provides binding energies very similar to the LN one. 
In particular,  it is surprising that, at the shell closures ($A = 100 $ and 132 ), 
the LN approach provides both, binding energies very close to the HF ones and, 
at the same time,  larger particle-particle correlation energies than the VAP method, 
see Fig.~\ref{Fig1}. These results  could be interpreted  again
as a degradation of the LN approach in the very weak pairing regime. The  PLN results
on the contrary become closer to the VAP ones.  

On the base of general arguments
one would expect larger similarity between the bulk behavior of $E_{pp}$ and 
$ \Delta E_G$ than the one found. A close look at the different solutions
reveals that while in all approaches (with the exception of the HF one) all nuclei under 
study remain spherical, in the HF one some nuclei get deformed due to the absence of 
pairing correlations.  That means in the upper panel
of Fig.~\ref{Fig2} some deformation effects are present and the comparison
of  $ \Delta E_G$ with $E_{pp}$ is not as obvious as if these effects
would not be there. To eliminate deformation
effects we have performed {\it spherical} HF  calculations. The quantities
$ \Delta E_G$ evaluated with the spherical binding energies are displayed
in the lower panel of Fig.~\ref{Fig2}. 
As expected, since the HF energies are now smaller -we restrict ourselves
to spherical shapes- we obtain larger values  for $ \Delta E_G$.  As before
PAV and LN values are close to each other and more interestingly the two hump 
structure found in Fig.~\ref{Fig2} is recovered. We observe that, with the 
exceptions of the shell closures, the different curves behave more similar
one to each other than they do in the particle-particle energy of Fig.~\ref{Fig1}.

\begin{figure}[h]
\begin{center}
\parbox[c]{15cm}{\includegraphics[width=14cm]{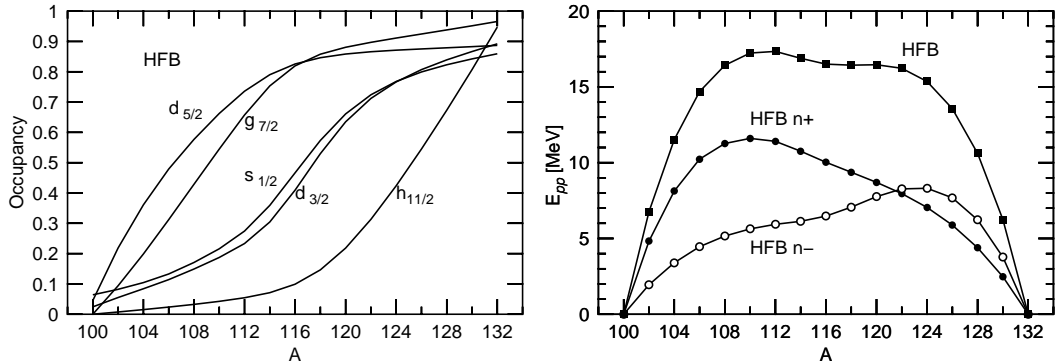}}
\end{center}
\caption{Left panel: Occupancies of the active orbitals in the HFB approach as
a function of the mass number. Right panel: The neutron particle-particle 
correlation energy  in the HFB approximation for positive and negative parity.}
\label{Fig3}
\end{figure}

To study the origin of the two hump structure we shall investigate the 
fractional harmonic oscillator shell occupancy, defined by
\begin{equation}
\nu (n,l,j) = \frac{1}{2j+1} \sum_{m=-j}^{j} 
\langle \Phi | c^{\dagger}_{nljm} c_{nljm} | \Phi \rangle.
\end{equation}
In the left part of Fig.~\ref{Fig3} we represent the occupancies of
the pairing active shells in the HFB approximation (for the other
approaches the conclusions do not change). As expected at $N=50$ the
shells are empty and at $N=82$ filled (the small deviations from  zero
and one are due to the fact that in the spherical HFB the quantum  
number $n$ is not conserved) and in between a smooth filling of the shells
takes place.
This smooth filling may give rise to divergences in the PNP approaches
when the occupancies $v_k^2$  take the value 0.5 and the exchange terms 
have been neglected \cite{AER.01P} in the calculations. 
The largest pairing correlations are expected from the big shells and 
at half shell occupancy. Accordingly, we expect a maximum around $A= 108-110$ 
from the d$_{\frac{5}{2}}$ and  g$_{\frac{7}{2}}$ orbitals and another one
 around $A= 124-126$
 stemming from the  h$_{\frac{11}{2}}$ orbit. Looking at the right panel of 
Fig.~\ref{Fig3}, where we have plotted separately the contribution 
to the total pairing energy of  positive (HFB n+) and negative (HFB n-) 
parity shells, we find the expected behavior. The deep in the pairing energy
around $A =116$ is due to the fact that the gain in pairing of the
small shells  s$_{\frac{1}{2}}$ and  
d$_{\frac{3}{2}}$, which  are being filled around this mass number, does not 
compensate the loss of pairing due to the higher occupancy of the  
d$_{\frac{5}{2}}$ and  g$_{\frac{7}{2}}$ shells.
A possible explanation of  the fact that the HFB, LN, PAV and PLN give 
values for $E_{pp}$ with a larger dispersion around the major midshell  than
at the beginning or at the end of the shell, might be that in the first case
one has to deal with five open subshells while in the second one only with
one or two.

Taking into account these results we may understand somewhat better the relation 
between $ \Delta E_G$ and  ${ E_{pp}}$. The total energy in a given approach 
can be written  as $E_{approach} = E_{HF} (\varphi_{approach}) + E_{pp}$,
where $E_{HF} (\varphi_{approach})$ represents the kinetic energy and 
the contribution of the Hartree-Fock field to the energy~\footnote{ In the LN 
case we have to add the term  $- h_2 <(\Delta \hat{N})^2>$.}calculated 
with the wave function $\varphi_{approach}$. This expression is also valid 
for the projected energy  (PLN, PAV and VAP), see Eq.~(B-1) of reference \cite{AER.01P}. 
Substitution of  $E_{approach}$ in Eq.~(\ref{eq:enepc}) provides
$  \Delta E_G  =  E_{HF} - E_{HF} (\varphi_{approach}) - E_{pp}$.
 The quantity $|E_{HF} - E_{HF} (\varphi_{approach})|$ represents the loss
of energy due to the readjustment of the plain HF occupancies caused by the
pairing field. Looking at Fig.~\ref{Fig3}, we find that near the shell closures
the occupancies, independently of the approach, are either zero or one, i.e., the
HF occupancies are not very different from the ones of the corresponding 
approach and we do 
not expect large differences between  $ \Delta E_G$ and  ${ E_{pp}}$. As we move 
from closed shells towards midshell the fractional occupancies increase deviating
from the plain HF ones. The largest deviations between $ \Delta E_G$ and  ${ E_{pp}}$
are expected, therefore, in the middle of the shell. A close look at Figs.~\ref{Fig1} 
and \ref{Fig2} confirms these expectations. An interesting point is the inversion of the
HFB (LN) and PAV (PLN) curves in Fig.~\ref{Fig2} as compared with Fig.~\ref{Fig1}.
The HFB (LN) line is below  the PAV (PLN) in Fig.~\ref{Fig2} at variance with 
Fig.~\ref{Fig1} because the projected occupancies (PAV or PLN) are closer to the plain HF,
i.e., smaller $|E_{HF} - E_{HF} (\varphi_{approach})|$, than the 
unprojected ones (HFB or LN).

\begin{figure}[h]
\begin{center}
\parbox[c]{10cm}{\includegraphics[width=8cm]{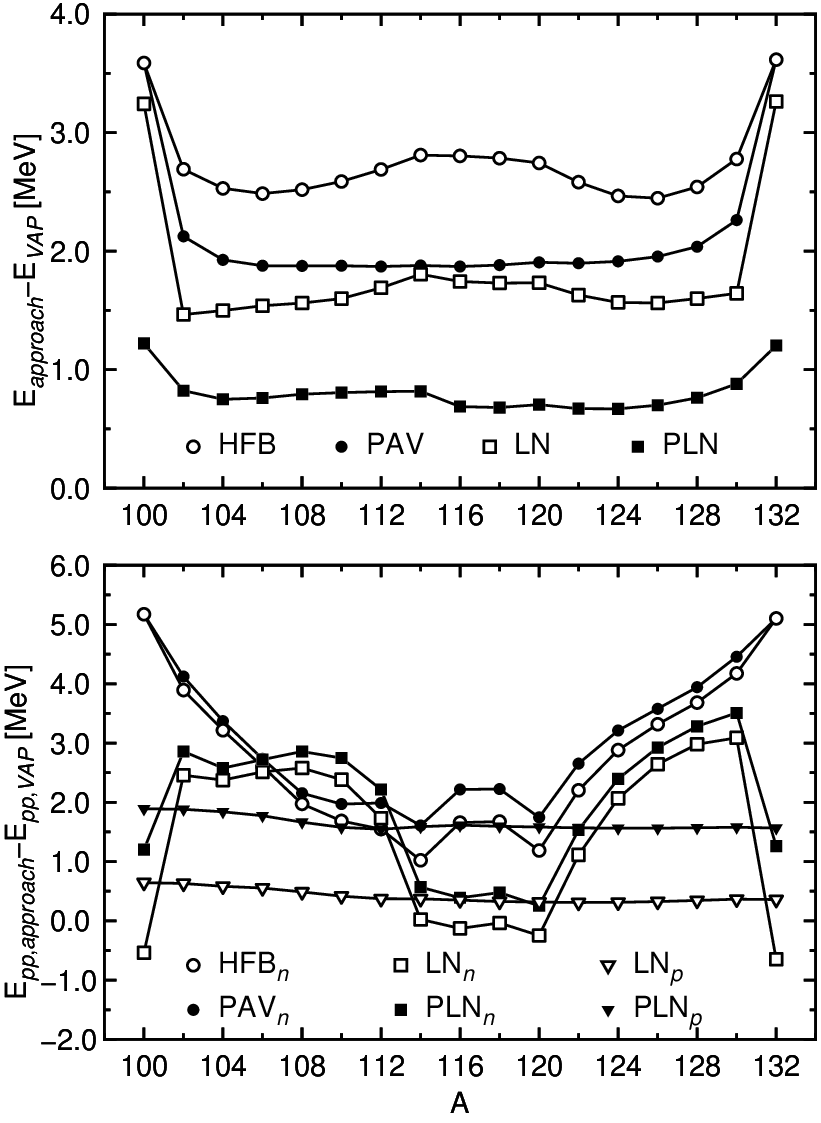}}
\end{center}
\caption{Upper part: The energy difference between the binding energy in the 
different  theoretical approaches and the VAP approach as a function of the
mass number. Lower part: Same as upper part for the particle-particle
correlation energy.}
\label{Fig4}
\end{figure}

The HFB, PAV, LN and PLN  approaches are approximations to the full
variation after projection method. It would be interesting to see how much
the different approximations deviate from the full VAP method which they try
to emulate. In the upper part of Fig.~\ref{Fig4} we plot the difference
of the binding energies calculated in the different approaches and the VAP one.
In the HFB case we find on the average deviations of about 2.6 MeV with
the exception of the nuclei in the  neighborhood of the shell closures
and around $A=116$ where we get larger values. These results are in
qualitative 
agreement with the Kamlah expansion, according to which the larger the
deformation in the gauge space associated with the symmetry operator the
better the expansion will be. In the right panel of Fig.~\ref{Fig1} we
have represented the fluctuations for the particle number operator, which
give a measure of the gauge deformation as a function of the mass number.
We find a qualitative correlation between $<(\Delta N )^2>$ and the goodness
of the HFB approach.
We find that for those mass numbers where $<(\Delta N )^2>$ has large values
 the HFB approach gets closer to the VAP  method.
 The PAV results provide, on the average, an additional lowering
of about 0.75 MeV. This approximation is rather uniform as a function of the
mass number and only the two or three nuclei close to the shell closures 
deviate more from the average. The  LN  values are on the average about 1.7 MeV
higher than the VAP and as the other approximations they differ at most from
the VAP at the shell closures. However, in the LN  {\it only} 
the $N=50$ and $N=82$ isotopes deviate strongly from the VAP values. 
A quantitative change is provided by the PLN. Its binding energies differ over a 
wide range by only about 0.75 MeV from the VAP, and even in 
the worst cases, for $N=50$ and $N=82$, the deviations is
only about 1.25 MeV.

The discussion of the upper part of Fig.~\ref{Fig4} does not give us 
information on the content of the wave functions, it tells us only about the 
ability of the different approaches to reproduce the binding energy of the
VAP method which they try to emulate. Since the total energy is the sum of
several terms, there is no guarantee that each term reproduces with the same
quality the corresponding term in the VAP approach.
To gain more insight into the wave function
we have plotted in the lower panel of  Fig.~\ref{Fig4} the difference
between the particle-particle correlation energy of the HFB, PAV, LN and PLN approaches
and the VAP method. Let us first concentrate on the neutron parts. The first
observation is that these quantities show a stronger dependence on  the
mass number than the binding energies. The
HFB and PAV results are close to each other, though the HFB ones get closer 
to the VAP than the PAV, contrary to what happened with the total 
energies. Furthermore they approach the VAP results best around $A=116$, also
at variance with the binding energy results. The LN and PLN results also behave
similar, they approximate  best the VAP results in the region around $A=116$
and at the shell closures, though the good agreement at the shell closures could
be fortuitous as we shall see below. Furthermore, the LN results get closer to
the VAP than the PLN. These features of the LN and PLN are also the contrary to
what we obtained for the total energies.

\begin{figure}[h]
\begin{center}
\parbox[c]{15cm}{\includegraphics[width=14cm]{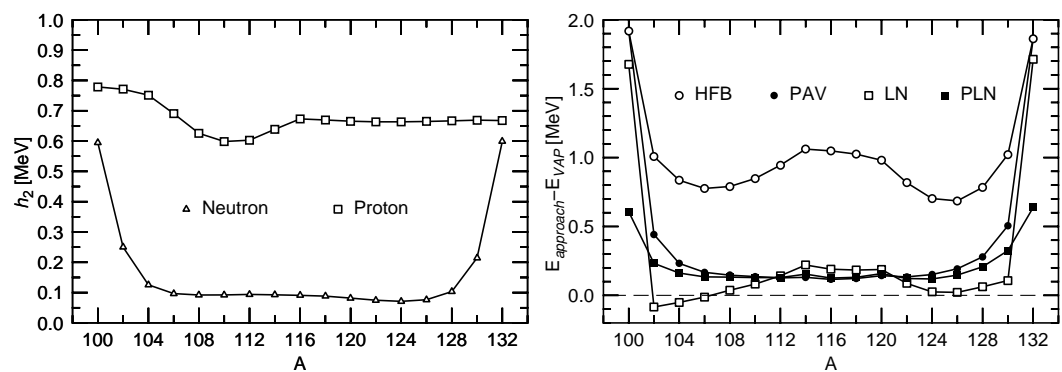}}
\end{center}
\caption{Left panel:  The $h_2$ coefficient of the LN approach
as a function of the mass number. Right panel: The energy difference between 
the binding energy of the HFB, PAV, LN or PLN approach and the VAP one,  for the solutions 
with no proton pairing correlation, as a function of the mass number. }
\label{Fig5}
\end{figure}

In  the lower panel of  Fig.~\ref{Fig4}, we see the discontinuities in the  LN and PLN 
approaches  at  A=102  and A=130, already  commented in reference to Fig.~\ref{Fig1}. 
 This behavior could be associated with a failure of the LN expansion at the
phase transitions.
As a matter of fact in reference \cite{VER.01} it has been shown that large changes 
in the  $h_2$ parameter are associated with the breakdown of
the second order expansion of the projected energy. On the left
hand side of Fig.~\ref{Fig5} we have plotted the $h_2$ parameters as a function 
of the mass number.
 As one can see the neutron $h_2$ parameter is rather constant and small from A=106 up 
to A=126, then it rises to the large shell closure value of 0.6. For protons we
find large and rather constant $h_2$  values (the small dip around A=110 has to do 
with the small bump at the same place in $<(\Delta N)^2>$, see Fig.~\ref{Fig1}). From this
plot we conclude that the origin of the discontinuity is the fast change in  $h_2$,
which itself is caused by the breakdown of the expansion at second order.
 The overshooting of  ${ E_{pp}}$ in the LN approach at $A=100$ and $A=132$
is caused by the term $-h_2 (\Delta \hat{N})^2$, in Eq.~(\ref{eq:ln})~: The large $h_2$ 
value makes that term very big leading to an exaggerated scattering of  neutrons pairs 
across the Fermi surface.  
This overshooting is equivalent to the behavior found in the two-level pairing
model where the LN energy, in the weak pairing regime, is deeper than the exact 
one\cite{ZSF.92}. 
  Concerning the proton particle-particle correlation energies we observe that the LN and 
PLN, as expected, stay rather constant with the mass number. The overshooting
of the LN approach does not appear in this case, probably, because of the Coulomb antipairing
effect \cite{AER.01}. 

 In the calculations we have chosen spherical nuclei in order to separate 
deformation effects from pairing effects, though in our triaxial codes one may get
solutions with very small  deformations.  Now we would like to furthermore separate 
proton and neutron pairing effects. 
  The comparison of the HFB and PAV binding energies with the VAP results for the
Sn isotopes (proton closed shell) is not very fair because no proton pairing 
is obtained in the HFB and PAV approximations whereas this is not the case
for VAP.  The comparison of the LN and PLN approaches is not fair either because, 
as we have seen, for closed shells the LN expansion is not a good one. That means,
the results for the proton channel may   overshadow the results of the neutron channel.
The comparison will be more equitable for ordinary nuclei
(proton and neutron shells open) where the HFB approach will provide pairing
in both channels and the LN one will perform better. In order to disentangle the different
contributions in the Sn isotopes we have also performed VAP and LN calculations with the additional
constraint~\footnote{This can  easily be
done by restricting the proton  intrinsic variational wave function to be of the HF 
(not HFB !) type.} that no proton pairing is allowed. In this way the proton channel is treated
in the same way in all approximations, i.e., the wave function is always of the HF type. 
In the right  panel of Fig.~\ref{Fig5} we display the difference of the
binding energy in the HFB, PAV, LN or PLN approach and the VAP one,  the LN (PLN) and
VAP energies calculated with the mentioned constraint. In the HFB, PAV, LN and PLN approaches, 
with the exception of the shell closures at  N=50 and N=82, we are able to reproduce
the VAP results, on the average,  within 1, 0.2, $\pm 0.2$ and 0.15 MeV, respectively. 
The projected versions PAV and PLN provide smoother approximations, i.e, less isotope dependence, 
than the unprojected ones. The LN approach provides in some cases slightly deeper
energies than the VAP one. In the shell closures we find deviations of about 2 MeV
in the HFB, PAV and LN approaches, only the PLN reproduces the total binding energy
within 0.6 MeV. Looking at this figures one would be tempted to say that the weak pairing
regime in the ground state\footnote{Excited states, specially high-spin ones, are 
explicitly excluded. } of atomic nuclei only take place for closed shell nuclei. We may now
return to the question on the quality of the LN  and PLN approaches in the weak pairing regime.
Let us compare the  LN and PLN  results of the right panel of Fig.~\ref{Fig5} from $A = 106$ to
$A = 124$, only strong pairing regime, with the corresponding ones in the upper panel of Fig.~\ref{Fig4},
weak (strong) pairing regime in the proton (neutron) channel. We find a certain degradation of the LN 
approach in the weak pairing regime, Fig.~\ref{Fig4}, since it deviates too much from the VAP as 
compared with Fig.~\ref{Fig4}.  The PLN approach, on the other hand, provides rather reasonable results
in the weak proton pairing regime. 
Concerning to the neutron pairing energies $E_{pp}$ in the VAP approximation, they are almost unchanged 
by the fact that proton pairing correlations are not allowed.  Therefore, the difference 
in the neutron  pairing energies in the HFB (PAV) and the VAP one is still given by the bottom panel
of Fig.~\ref{Fig4}.

In conclusion, we have investigated the behavior of the pairing correlations
along a major shell in the variation after projection method plus four  approximations to it
with the effective Gogny force.  
If we look at the total binding energies we find that the best approximation to 
the VAP is provided by the PLN, followed by the LN or PAV and  HFB. The PAV results being 
surprisingly good in midshell. The crucial test of the 
goodness of the approaches is provided by the shell closures and their nearest
neighbors. There, only the PLN provides a reasonable approximation to the 
VAP method. For the other isotopes, all approaches are rather uniform and the 
quality of a given  approach is almost independent of the mass number. We  predict that the best 
approximations will be obtained for doubly open shell nuclei. 
We also show that the weak pairing regime in the ground state
of atomic nuclei is limited to the nearest neighborhood of closed shells. The isotope dependence
of the binding energies is smoother in the PAV and PLN approaches than in the HFB and LN ones,
indicating that this behavior is caused by admixture of wrong particle number 
components in the wave function of the HFB and LN approaches.
On the other hand, if we look at the wave function content, on the pairing
correlation energies for example, we find that the results are strongly approach
dependent and, in general, less uniform than the ones for binding energies.
The ability of the different approximations to reproduce the particle-particle correlation
energy of the VAP approach is worse than the one to reproduce the total binding energy. The
last statement is obvious from the variational point of view.  
 Specifically, we have found that mean field based theories like  HFB are able 
to provide the total binding energies of doubly open shell nuclei up to 2 MeV 
(1 MeV per channel) of the  VAP approach. For the LN, PAV and PLN theories the accuracy is much
better (0.2 MeV per channel). However, the predictions for the 
pairing correlation energy in  the HFB or PAV (LN or PLN) approach are about 3 MeV (2 MeV) 
away from the  VAP approach, which corresponds to a 20 per cent (15 per cent) 
accuracy. Consequently, properties strongly dependent on the pairing correlations,
like moments of inertia, level densities or excitation energies of the excited states, among others,
may not be well described in non-VAP theories in spite of providing high accuracy in the 
total binding energy as compared with the VAP prediction.
 
We would like to remark, lastly, that these conclusions are based on the mean field theories most 
widely used in nuclear structure calculations. In  principle one should compare these approaches, 
even the VAP one, with theories beyond mean field, which explicitly take into account additional 
pairing fluctuations.

\begin{ack}
This work was supported in part by DGI, Ministerio de Ciencia y Tecnologia, Spain, 
under Project BFM2001-0184.
\end{ack}


\end{document}